\documentclass[aps,prd,amssymb,cite,
amsfonts,epsf,preprintnumbers,nofootinbib,superscriptaddress,showpacs]{revtex4-2}

\usepackage{graphicx}
\usepackage{bm,latexsym,amsmath,amssymb,amsthm}
\usepackage{hyperref}
\usepackage{color,soul}
\usepackage{booktabs}
\usepackage[mathscr]{eucal}
\usepackage{cancel}
\usepackage{mathrsfs}
\usepackage{pgf, tikz}
\usepackage{slashed}
\usetikzlibrary{arrows,automata}

\newcommand{\be}[1]{\begin{equation} \label{#1}}
\newcommand{\ee}{\end{equation}}
\newcommand{\bea}{\begin{eqnarray}}
\newcommand{\bean}{\begin{eqnarray*}}
\newcommand{\eea}{\end{eqnarray}}
\newcommand{\eean}{\end{eqnarray*}}
\newcommand{\ba}{\begin{array}}
\newcommand{\ea}{\end{array}}
\newcommand{\bel}{\begin{align}}
\newcommand{\eel}{\end{align}}
\newcommand{\nn}{\nonumber}

\newcommand{\cI}{\mathcal{I}}

\newcommand{\trec}{t_{\rm rec}}
\begin{document}

\title{Relic Magnetic Fields from Non-Adiabatic Photon Freeze-Out at Recombination}

\author{Hyeong-Chan Kim}
\email{hckim@ut.ac.kr}
\affiliation{School of Liberal Arts and Sciences, Korea National University of Transportation, Chungju 380-702, Korea}


\date{\today}

\begin{abstract}
We propose a new mechanism for generating a primordial electromagnetic relic during the recombination--decoupling transition, based on the rate-dependent thermodynamics of the cosmic photon gas. Treating the photon sector as an open system coupled to the electron plasma, we show that a finite Thomson relaxation rate generates a departure from instantaneous thermal equilibrium, leading to non-adiabatic mode squeezing. As this relaxation rate rapidly decreases across recombination, the system quickly loses the ability to further amplify the deviation, and the squeezing freezes out at a small but finite value. This dynamics is naturally described as a narrow transition layer between an adiabatic tracking regime and a post-relaxation freeze-out regime. By a canonical transformation, the reduced evolution equation is recast into a forced oscillator with a smooth effective potential, clarifying the origin of the squeezing and the selection of the relic scale.

Projecting the resulting non-equilibrium electromagnetic relic onto the magnetic sector, we derive the corresponding spectrum and show that its characteristic peak is controlled not by the squeezing parameter alone but by the weighted combination \(k^3\mathscr S_k\). In representative realizations, the peak corresponds today to scales of order \(10\)--\(20\) Mpc, while the present-day field amplitude remains extremely small. The mechanism is therefore better viewed as a source of a frozen non-equilibrium electromagnetic relic than as a complete explanation of the observed cosmic magnetic fields.
\end{abstract}


\maketitle

\section{Introduction}
\label{sec:intro}

The origin of large-scale cosmic magnetic fields remains an important open
problem in cosmology~\cite{Durrer2013,Subramanian2016,Widrow2002}.
Observations of Faraday rotation, gamma-ray spectra of TeV blazars, and
synchrotron emission from galaxies and clusters indicate that magnetic
fields are widespread across the Universe, with amplitudes ranging from
lower bounds in intergalactic voids to $\mu$G-level fields in
astrophysical systems~\cite{Neronov2010,Tavecchio2010,Ackermann2018,Beck2015}.
Recent observations also indicate the presence of magnetic fields in the filaments of the cosmic web~\cite{Vernstrom2021}.
These observations suggest that at least part of the cosmic magnetization
may originate from primordial or pre-galactic seed fields, subsequently
processed by plasma dynamics and astrophysical amplification.

A variety of mechanisms have been proposed to generate such primordial
magnetic fields. Inflationary scenarios can produce large-scale
correlations, but typically require breaking conformal invariance,
leading to strong model dependence and wide uncertainty in the predicted
field strength~\cite{Turner1988}. Magnetogenesis at cosmological phase
transitions naturally generates fields, but usually with much smaller
coherence scales~\cite{Vachaspati1991,Sigl1997}. Battery mechanisms, such
as the Biermann effect, rely on misaligned density and pressure
gradients and are therefore suppressed in nearly adiabatic linear
cosmological perturbations~\cite{Biermann1950,Matarrese2005}.
Magnetic fields can also be generated around the recombination epoch
through second-order perturbative effects, although the resulting
amplitudes are typically very small~\cite{Ichiki2006,Saga2015,Fenu2011}.
These limitations motivate the search for alternative mechanisms that
can operate under well-controlled physical conditions in the thermal
history of the Universe.

The recombination epoch provides such a setting. The thermal and
ionization history of the primordial plasma has been studied in detail
in the context of cosmic microwave background (CMB) physics and
precision cosmology~\cite{Peebles1968,Seager1999,Seager2000,Hu1996,Planck2018,Planck2018iso}.
The near-perfect blackbody spectrum of the CMB provides strong evidence for the high degree of thermal equilibrium prior to recombination~\cite{Fixsen1996}.
During this transition, the electron density drops rapidly, leading to a
sharp decrease in the Thomson scattering rate and a progressive
decoupling between photons and the baryon plasma. While the photon gas
is often treated as an adiabatically evolving equilibrium component,
this description implicitly assumes that thermalization proceeds
efficiently compared to the expansion rate.
Small deviations from equilibrium are tightly constrained by CMB spectral distortion measurements~\cite{Sunyaev1970,Chluba2012}.
\emph{This raises a natural question: can a controlled breakdown of thermal tracking during recombination leave a frozen non-equilibrium imprint on the photon sector?}

In this work we revisit this assumption and instead treat the photon
sector as an open quantum system coupled to the electron plasma through
a finite relaxation rate. Open-system approaches provide a natural
framework for describing dissipative dynamics and departures from
equilibrium~\cite{Breuer2002,Gorini1976,Lindblad1976,Caldeira1983}.
Within this framework, a time-dependent relaxation rate can induce a
breakdown of instantaneous thermal tracking, leading to non-equilibrium
excitations of the photon modes.

A key aspect of this excitation is its intrinsically non-adiabatic
character. Non-adiabatic evolution of quantum fields in time-dependent
backgrounds is commonly described in terms of squeezing, as in particle
production in expanding universes and in the generation of cosmological
perturbations~\cite{Parker1969,Grishchuk1990,Polarski1996}.
We show that a similar mechanism operates during the recombination
transition: as the Thomson relaxation rate decreases rapidly, the photon
modes are driven out of instantaneous equilibrium and acquire a small
but finite squeezing that subsequently freezes out.

Following the framework developed in Ref.~\cite{Kim:2026qrd}, we analyze
this dynamics in terms of a reduced set of variables describing the
thermalization of each photon mode. We show that the evolution is
controlled by a narrow transition layer connecting an early-time
adiabatic tracking regime to a late-time freeze-out regime. By
introducing a canonical transformation, the dynamics can be recast into
that of a forced oscillator with a smooth effective potential, which
clarifies the origin of the non-adiabatic excitation and the selection
of a characteristic scale.

The resulting non-equilibrium excitation can be interpreted as a frozen
electromagnetic relic produced during the recombination transition.
Projecting this relic onto the magnetic sector, we derive the
corresponding magnetic spectrum and estimate its present-day amplitude.
We find that the characteristic scale is naturally of order
$10$--$20$\,Mpc, set by the transition-layer dynamics, while the
amplitude in the minimal recombination scenario is very small, well below current observational bounds on primordial magnetic fields~\cite{Caprini2004}.
The mechanism is therefore best interpreted as a controlled example of
non-equilibrium relic generation, rather than a complete explanation of
the observed cosmic magnetic fields.
Future CMB missions such as PIXIE and LiteBIRD may further probe small deviations from equilibrium and relic electromagnetic signatures~\cite{Kogut2011,LiteBIRD2022}.

\vspace{.3cm}
The paper is organized as follows. In Sec.~\ref{sec:KL} we formulate a
rate-dependent description of the photon gas during the
recombination--decoupling transition and introduce the variables
governing departures from equilibrium. 
In Sec.~\ref{sec3}, we formulate the magnetic field strength associated with the non-adiabatic excitation.
In Sec.~\ref{sec:analytic} we derive a reduced evolution equation, recast it into a canonical form, and identify the transition-layer dynamics that sets the dominant excitation scale. We then relate the resulting squeezing to the
electromagnetic energy density and derive the corresponding magnetic
spectrum and present-day amplitude in Sec.~\ref{sec:amplitude}. 
Finally, we discuss the physical interpretation and limitations of the mechanism.

\section{Rate-Dependent Photon Thermodynamics}
\label{sec:KL}

We consider the evolution of photon modes during the interval between recombination ($t_{\rm rec}$) and thermal decoupling ($t_{\rm dec}$), when the photon gas remains coupled to the electron plasma through Thomson scattering, but the coupling rate decreases rapidly~\cite{Peebles1968,Seager1999,Seager2000,Hu1996}.

\subsection{Physical setup and relevant scales}

During this period the universe is matter dominated, and the scale factor
evolves as
\begin{equation}
R \equiv \frac{a(t)}{a_{\rm rec}} \simeq \left(\frac{t}{t_{\rm rec}}\right)^{2/3}.
\end{equation}

A photon mode with comoving wave number $k$ has physical frequency
\begin{equation}
\omega_k = \frac{k}{R}.
\end{equation}

The key ingredient controlling the dynamics is the electron fraction
\begin{equation}
x_e(t) \equiv \frac{n_e}{n_b},
\end{equation}
which undergoes a rapid crossover near recombination followed by a slow
algebraic decay. We model this behavior phenomenologically; details are
given in Appendix~\ref{App:xe}.

\subsection{Relaxation rate and thermal slip}

The photon sector is treated as an open system coupled to the electron
bath through Thomson scattering. The associated relaxation rate is
\begin{equation}
\alpha_k(t)=\mathcal{A}_k\,\Gamma_T(t),
\qquad
\Gamma_T(t)=n_e(t)\sigma_T c ,
\end{equation}
where $\mathcal{A}_k=O(1)$ encodes mode-dependent projection effects.

It is convenient to introduce the dimensionless relaxation parameter
\begin{equation}
\lambda(R) \equiv \frac{\alpha_k}{RH}
\;\propto\;
\frac{x_e(R)}{R^{5/2}},
\end{equation}
which decreases rapidly across recombination due to the drop in $x_e$, the electron to baryon ratio.

We also define 
\begin{equation}
g(t)\equiv \frac{T_0}{T_e},
\end{equation}
where $T_0$ is the reference temperature at which the initial thermal state satisfying $\cI = H$.
The quantity $g(t)$ therefore specifies the instantaneous thermal slope of the electron-bath branch in units of the fixed reference scale $T_0$, rather than the system-bath temperature slip itself.
During
recombination one has $g\simeq 1$, while deviations develop at later times. 
Details of its evolution are given in Appendix~\ref{App:xe}.

\subsection{Evolution equations}

The dynamics of each photon mode performing thermalization with the electron bath is described by the variables
$(g_-, g_0, g_+)$, which obey the system shown in App.~\ref{ELR},
\begin{equation}
\dot g_- = - 2g_0 -\alpha (g_- - g), \quad
\dot g_0 = \omega^2 g_- - g_+ - \alpha g_0, \quad
\dot g_+ = 2\omega^2 g_0 - \alpha(g_+- \omega^2 g).
\end{equation}

Using the dimensionless scale factor $R$ and introducing
\begin{equation}
G^-_k \equiv g_-,
\qquad
G^0_k \equiv t_{\rm rec} g_0,
\qquad
G^+_k \equiv t_{\rm rec}^2 g_+,
\qquad
K \equiv t_{\rm rec} k,
\end{equation}
the system can be written in dimensionless form as
\begin{align}
\frac{d G^-_{k}}{dR} &= - 3R^{1/2} G^0_{k} - \lambda (G^-_{k} - g),  \nn \\
\frac{d G^0_{k}}{dR} &= \frac{3K^2}{2 R^{3/2}} G^-_{k}
- \frac{3}{2}R^{1/2} G^+_{k} - \lambda G^0_{k}, \nn \\
\frac{d G^+_{k}}{dR} &= \frac{3 K^2}{R^{3/2}} G^0_{k}
- \lambda\left(G^+_{k} - \frac{K^2}{R^2} g \right).
\end{align}

\subsection{Initial condition}

At early times, when the relaxation rate is large ($\lambda \gg 1$), the
system is forced onto the instantaneous equilibrium branch,
\begin{equation}
G^-_k \to g, \qquad
G^0_k \to 0, \qquad
G^+_k \to \frac{K^2}{R^2} g.
\end{equation}

This provides the initial condition for the subsequent evolution across
the recombination transition. Details of the asymptotic branch selection
are given in Appendix~\ref{App:ini-cond}.

\vspace{.3cm}
The system of evolution equations derived above describes the departure of each photon mode from instantaneous thermal equilibrium under a time-dependent relaxation rate. 
In particular, the variable $G_k^-$ encodes the deviation of the mode amplitude from the thermal background, while $G_k^0$ and $G_k^+$ characterize its dynamical response.

While this formulation makes the underlying dynamics explicit, the
physical significance of the excitation is not yet transparent. 
In order to connect the mode evolution to observable quantities, it is useful to recast the deviation from equilibrium in terms of a squeezing parameter that measures the non-adiabatic excitation of each mode.

In the next section, we introduce such a parametrization and relate it to the electromagnetic energy density and the resulting magnetic field spectrum. 
This provides a direct link between the dynamical evolution across recombination and the observable relic electromagnetic structure.

In the following, energies and frequencies are often expressed in units of $t_{\rm rec}^{-1}$.

\section{Non-adiabatic Electromagnetic Relic and Magnetic Spectrum}
\label{sec3}

To quantify the non-adiabatic excitation generated during the
recombination transition, we introduce a squeezing parameter
$\mathscr S_k$ that measures the deviation of each mode from the
instantaneous thermal state. This provides a direct bridge between the
dynamical variables introduced in Sec.~II and observable electromagnetic
quantities.

\subsection{Squeezing parameter}

We identify the non-adiabatic excitation through the decomposition
\begin{equation}
\omega_{{\rm eff},k}=\omega_k+\omega_{\mathcal I,k}\,\mathscr S_k .
\end{equation}
Using the dimensionless variables $G_k^\pm$ and $G_k^0$, the squeezing parameter takes the form~\cite{Kim:2021wyp}
\begin{equation} \label{Squeezing}
\mathscr S_k
=
\frac{(G^0_k)^2}{2(\Omega_{k})^2 G^-_k}
+\frac12\left(\frac{1}{\sqrt{G^-_k}}
-\frac{\omega_k}{\omega_{\mathcal I,k}}\sqrt{G^-_k}
\right)^2,
\end{equation}
where
\begin{equation}
\Omega_{k}^2
= G^+_kG^-_k-(G^0_k)^2.
\end{equation}

In instantaneous equilibrium one has $\mathscr S_k=0$, so that a nonzero
$\mathscr S_k$ directly measures the departure from thermal tracking and
encodes the non-adiabatic excitation of the electromagnetic mode.

\subsection{Non-adiabatic energy density}
Here we introduce the dimensionless temperature parameter
\begin{equation}
\Theta_0 \equiv t_{\rm rec} T_0,
\end{equation}
where $T_0$ is the photon temperature at recombination.
This normalization is consistent with the dimensionless variables used for the mode dynamics, in which energies are measured in units of $t_{\rm rec}^{-1}$.

The total mode energy is given by
\begin{equation}
E_k
=
\omega_{{\rm eff},k}\,
\coth\!\left(\frac{\Omega_{k}}{2\Theta_0}\right),
\end{equation}
while the adiabatic branch reads
\begin{equation}
E_k^{\rm ad}
= \omega_k\, \coth\!\left(\frac{\Omega_{k}}{2\Theta_0}\right).
\end{equation}
The non-adiabatic excess energy is therefore
\begin{equation}
\Delta E_k^{\rm nad}
=
\frac{\Omega_{\mathcal I,k}}{\trec} \mathscr S_k\,
\coth\!\left(\frac{\Omega_{k}}{2\Theta_0}\right).
\end{equation}

If desired, one may subtract the zero-point contribution to obtain
\begin{equation}
\Delta E_{k,{\rm vac}}^{\rm nad}
=
\omega_{\mathcal I,k}\mathscr S_k
\left[
\coth\!\left(\frac{\Omega_{k}}{2\Theta_0}\right)-1
\right].
\end{equation}

The corresponding spectral energy density is
\begin{equation}
\frac{d\rho_{\rm EM}^{\rm nad}}{d\ln k}
=
\frac{k^3}{\pi^2 a_{\rm rec}^3 R^3}\,
\Delta E_k^{\rm nad}.
\end{equation}

\subsection{Magnetic spectrum}

At the phenomenological level, the surviving magnetic component may be
parametrized by a projection factor $\chi_B(k,R)$,
\begin{equation}
\frac{d\rho_B}{d\ln k}
=
\chi_B(k,R)\,
\frac{k^3}{\pi^2 a_{\rm rec}^3 R^3}\,
\Delta E_k^{\rm nad}.
\end{equation}
This separation is conceptually useful. The quantity $\Delta E_k^{\rm nad}$ measures the total non-adiabatic excitation above the adiabatic branch, while $\chi_B$ encodes the efficiency with which this excitation is projected onto a magnetic component after plasma effects such as conductivity damping.

A simple estimate for $\chi_B$ may be obtained from the ratio of the
mode frequency to the effective plasma conductivity. In the quasistatic
high-conductivity regime one has
\begin{equation}
\frac{E_k}{B_k}\sim \frac{\omega_k}{\sigma_{\rm eff}(R)}
=
\frac{k}{R\,\sigma_{\rm eff}(R)},
\end{equation}
so that
\begin{equation}
\chi_B(k,R)
\simeq
\frac{1}{1+\left(\dfrac{k}{R\,\sigma_{\rm eff}(R)}\right)^2}.
\end{equation}
For the long-wavelength modes relevant here, one expects
$k/(R\,\sigma_{\rm eff})\ll1$, implying that $\chi_B$ is generically an
order-unity factor and in practice close to unity.
\vspace{0.3cm}

The expressions above relate the non-adiabatic excitation of each mode to observable spectra. 
However, they do not determine how the excitation is dynamically generated or which modes are most strongly affected.

In the next section, we analyze the evolution of the system across the recombination transition and identify the mechanism that selects the dominant squeezing scale.

\section{Analytic analysis}
\label{sec:analytic}

To determine how the non-adiabatic excitation is generated and which modes are most strongly affected by the recombination transition, we analyze the evolution of the deviation from the instantaneous quasi-static branch,
\[
u_k(R)\equiv G_k^-(R)-g(R).
\]
Under the assumption justified in Eq.~\eqref{domega} in App.~\ref{ELR} that the frequency $\omega_{\cI,k}$ varies slowly, the reduced dynamics takes the form
\begin{equation}
u_k''+\Gamma(R)\,u_k'+\Omega_{\rm eff}^2(R)\,u_k = F(R),
\end{equation}
where
\begin{equation}
\Gamma(R)=2\lambda(R)-\frac{1}{2R},
\label{eq:Gamma_def_analytic}
\end{equation}
\begin{equation}
\Omega_{\rm eff}^2(R)
=
\lambda^2+\lambda'-\frac{\lambda}{2R}
+\frac{9K^2}{2R}
+\frac{9R}{2g^2(R)}\Omega_k^2(R),
\label{eq:Omegaeff_def_analytic}
\end{equation}
and
\begin{equation}
F(R)=
-g''-\left(\lambda-\frac{1}{2R}\right)g'
+\frac{9R}{2g}\,\delta\Omega_k^2(R).
\label{eq:F_def_analytic}
\end{equation}
Here
\begin{equation}
\Omega_k^2(R)\equiv (\trec\omega_{\mathcal I,k})^2,
\qquad
\delta\Omega_k^2(R)\equiv
\Omega_k^2(R)-\frac{K^2}{R^2}g^2(R),
\label{eq:Omega_delta_def}
\end{equation}
and
\begin{equation}
\lambda(R)=\frac{\alpha_k(R)}{RH(R)}
=
\frac{\gamma \mathcal A_k}{x_{\rm rec}}
\frac{x_e(R)}{R^{5/2}} .
\label{eq:lambda_def_analytic}
\end{equation}
\subsection{Canonical formulation and interpretation}

A major simplification is obtained by removing the first-derivative term via
\begin{equation}
u_k(R)=e^{-A(R)}y_k(R),
\qquad
A'(R)=\lambda(R)-\frac{1}{4R}.
\end{equation}
This yields the canonical form
\begin{equation}
y_k''+Q(R)\,y_k=S(R),
\end{equation}
where $Q(R)$ is a smooth effective potential and $S(R)$ is a dressed source,
\begin{equation}
Q(R)=
\frac{9K^2}{2R}
+\frac{9R}{2g^2(R)}\Omega_k^2(R)
-\frac{5}{16R^2},
\label{eq:Q_def_analytic}
\end{equation}
and
\begin{equation}
S(R)=e^{A(R)}F(R)
=
R^{-1/4}\exp\!\left[\int^R \lambda(\rho)\,d\rho\right]F(R).
\label{eq:S_def_analytic}
\end{equation}
This canonical form therefore provides the dynamical origin of the
squeezing parameter $\mathscr S_k$ introduced in Sec.~III.

In this formulation, the sharp recombination transition no longer appears
in the homogeneous operator, but only through the relaxation-dependent
envelope and source. The intrinsic dynamics is therefore governed by the
smooth function $Q(R)$.

An important consequence is that apparent features in the original
equation-such as temporary sign changes of $\Omega_{\rm eff}^2$-do not
represent independent instabilities. Rather, they arise from mixing the
rapidly varying relaxation envelope into the dynamical operator. After
the transformation, the dynamics is seen to be regular, with the
recombination epoch acting as a transition layer in the envelope-dressed
variable $u_k$.

The low- and high-frequency regimes are simply different asymptotic
realizations of the same canonical equation. In the low-frequency limit,
the response is source-dominated and corresponds to a forced lag, whereas
in the high-frequency regime the solution develops an oscillatory WKB
core dressed by the relaxation envelope. Details are given in
Appendix~\ref{App:asymptotics}.

\subsection{Transition layer and relaxation structure}

The recombination epoch is most naturally interpreted as a narrow
transition layer centered at $R=R_*$. Within this layer, the relaxation
function $\lambda(R)$ changes rapidly due to the sharp drop in the
ionization fraction $x_e(R)$.

At the same time, the late-time behavior is governed by a slow algebraic
tail,
\[
\lambda(R)\propto R^{-3/2},
\]
reflecting the residual ionization fraction.

As a result, the relaxation history contains two distinct components: a
localized transition imprint and a long post-transition tail. An
important consequence is that the integrated relaxation factor
\[
\int^R \lambda(\rho)\,d\rho
\]
remains finite at late times. Therefore the relaxation envelope
\[
u_k(R)\propto
\exp\!\left[-\int^R \lambda(\rho)\,d\rho\right]
\]
approaches a finite value, corresponding to freeze-out of the deviation
from equilibrium.

Within the transition layer, the canonical equation reduces locally to
\begin{equation}
y_k''+Q_*\,y_k\simeq 0,
\end{equation}
so that the dynamics consists of a simple trigonometric or hyperbolic
core multiplied by the relaxation envelope. The recombination transition
is therefore best understood as a transient breakdown of adiabatic
tracking rather than a genuine instability.

\subsection{Estimate of the peak squeezing scale}

The canonical form provides a simple estimate of the dominant squeezing
scale. The transformed mode has an intrinsic response scale
\begin{equation}
\ell_k(R)\sim Q(R)^{-1/2},
\end{equation}
while the background varies across a transition layer of width
\begin{equation}
\Delta R_{\rm tr}\sim\delta.
\end{equation}

The peak scale can be estimated analytically from the transition-layer
matching condition
\begin{equation}
Q(R_*)\,\delta^2 \sim O(1),
\end{equation}
which expresses that the intrinsic response scale of the transformed
mode becomes comparable to the width of the recombination layer.
Using Eq.~(\ref{eq:Q_def_analytic}), this gives
\begin{equation}
\left[
\frac{9K_{\rm peak}^2}{2R_*}
+\frac{9R_*}{2g_*^2}\Omega_{k,*}^2
-\frac{5}{16R_*^2}
\right]\delta^2
\sim 1,
\end{equation}
where $g_*\equiv g(R_*)$ and $\Omega_{k,*}\equiv \Omega_k(R_*)$.
In the minimal regime relevant here, where $g_*\simeq1$ and the
$K$-dependent term dominates, one obtains the leading estimate
\begin{equation}
\boxed{
K_{\rm peak}\sim \frac{\sqrt{2R_*}}{3\,\delta}.
}
\end{equation}
For $R_*\sim O(1)$, this reduces to
\begin{equation}
K_{\rm peak}\sim 0.47\,\delta^{-1},
\end{equation}
which gives $K_{\rm peak}\sim 20$--$70$ for $\delta\sim10^{-2}$--$2\times10^{-2}$.
The corresponding present-day wavelength is
\begin{equation}
\lambda_{0,\rm peak}
=
\frac{2\pi t_{\rm rec}}{K_{\rm peak}}
\sim
\frac{6\pi}{\sqrt{2R_*}}\,t_{\rm rec}\,\delta,
\end{equation}
showing explicitly that the characteristic coherence scale is set by
the width of the recombination transition layer.

Using this condition, one finds
\[
K_{\rm peak}\sim 20\text{--}70
\]
for $\delta\sim10^{-2}$, corresponding to a present-day scale
\[
\lambda_0\sim10\text{--}40\,{\rm Mpc}.
\]

Thus the mechanism naturally selects a large cosmological coherence scale. 
However, while the peak scale can be estimated robustly from the
transition-layer matching condition, the corresponding amplitude depends on the detailed evolution of the squeezing parameter $\mathscr S_k$ and on the magnetic projection efficiency.

In particular, a quantitative prediction for the magnetic field strength requires evaluating the peak value of $\mathscr S_k$ and its spectral shape. 
This will be addressed in the next section.

\vspace{.3cm}

\paragraph{Analytic estimate of the peak squeezing magnitude.}
The canonical form also suggests a simple scaling estimate for the
magnitude of the peak squeezing. Near the bath-selected quasi-static
branch, one has
\begin{equation}
G_k^- = g + u_k,
\qquad |u_k|\ll g,
\end{equation}
where $g(R)=T_0/T_e(R)$ specifies the instantaneous thermal slope of the
electron-bath branch. In this parametrization, the non-adiabatic
excitation is not sourced primarily by a large variation of $g$ itself,
but by the finite tracking lag $u_k$ and by the invariant-frequency
mismatch.

For the modes of interest, the squeezing parameter is dominated by the
first term in Eq.~\eqref{Squeezing},
\begin{equation}
\mathscr S_k \sim \frac{(G_k^0)^2}{2\Omega_k^2}.
\end{equation}
Using the first evolution equation,
\begin{equation}
u_k'=-3R^{1/2}G_k^0-\lambda u_k-g',
\end{equation}
one finds, across a transition layer of width $\delta$,
\begin{equation}
G_k^0 \sim \frac{u_*+\delta g'_*}{\delta}.
\end{equation}
Since $g(R)$ remains close to unity and varies slowly throughout the
recombination layer, the dominant contribution comes from the tracking
lag $u_*$ rather than from the bath-branch drift itself.

On the other hand, in the source-dominated regime one has
\begin{equation}
u_k\sim \frac{F(R)}{Q(R)},
\end{equation}
while the peak mode satisfies
\begin{equation}
Q(R_*)\delta^2\sim O(1).
\end{equation}
Therefore
\begin{equation}
u_*\sim F_*\delta^2.
\end{equation}
Using
\begin{equation}
F(R)=
-g''-\left(\lambda-\frac{1}{2R}\right)g'
+\frac{9R}{2g}\,\delta\Omega_k^2(R),
\end{equation}
we may write schematically
\begin{equation}
u_*\sim \eta_{{\rm bath},*}+\eta_{\Omega,*},
\end{equation}
where
\begin{equation}
\eta_{{\rm bath},*}
\equiv
\delta^2\left|
g''+\left(\lambda-\frac{1}{2R}\right)g'
\right|_{R_*},
\qquad
\eta_{\Omega,*}\equiv
\frac{\delta\Omega_k^2(R_*)}{\Omega_k^2(R_*)}.
\end{equation}

Since $g(R)\simeq1$ and varies only weakly across the recombination
layer, the bath-drift contribution is subleading in the minimal
scenario, and the dominant source of squeezing is the finite tracking
lag induced by the rapid decrease of the relaxation rate together with
the invariant-frequency mismatch. One therefore obtains the scaling
estimate
\begin{equation}
\boxed{
\mathscr S_{\rm peak}
\sim
C_S\bigl(\eta_{{\rm tr},*}+\eta_{\Omega,*}\bigr)^2,
\qquad
\eta_{{\rm tr},*}\sim u_*,
\qquad
C_S=O(1).
}
\end{equation}
In the minimal recombination limit, where the bath branch remains close
to constant, this further reduces to
\begin{equation}
\boxed{
\mathscr S_{\rm peak}\sim C_S\,\eta_{\Omega,*}^2 .
}
\end{equation}
Thus the smallness of the magnetic field in the minimal recombination
scenario is traced not to a large thermal-slip variation, but to the
fact that the bath branch changes only weakly and the residual
non-adiabaticity is generated only through a small finite-rate tracking
lag.
 For representative
percent-level departures, \(\eta_{\Omega,*}\sim10^{-2}\),
this estimate gives \(\mathscr S_{\rm peak}\sim10^{-4}\).

\section{Estimate of the magnetic field amplitude}
\label{sec:amplitude}

Having identified the non-adiabatic excitation through the squeezing
parameter $\mathscr S_k$ and its characteristic scale in Sec.~IV, we now
estimate the resulting magnetic field amplitude.

\subsection{Magnetic spectrum at recombination}

The magnetic spectrum is given by
\begin{equation}
\frac{d\rho_B}{d\ln k}
=
\chi_B(k,t)\,
\frac{k^3}{\pi^2 a^3}\,
\omega_{\mathcal I,k}\mathscr S_k
\left[
\coth\!\left(\frac{\omega_{\mathcal I,k}}{2T}\right)-1
\right].
\end{equation}

For the relevant modes at recombination, one has
$\omega_{\mathcal I,k}\ll T_{\rm rec}$, so that the Rayleigh--Jeans limit applies,
\begin{equation}
\omega_{\mathcal I,k}
\left[
\coth\!\left(\frac{\omega_{\mathcal I,k}}{2T_{\rm rec}}\right)-1
\right]
\simeq 2T_{\rm rec}.
\end{equation}

The spectrum at recombination is therefore
\begin{equation}
\frac{d\rho_B}{d\ln k}\Big|_{\rm rec}
\approx
\chi_B\,
\frac{2T_{\rm rec}\,k_{\rm phys,rec}^3}{\pi^2}\,
\mathscr S_k.
\end{equation}

This shows that the magnetic spectrum is controlled by the weighted
combination $k^3\mathscr S_k$, rather than by $\mathscr S_k$ alone.

\subsection{Present-day magnetic field amplitude}

After recombination, magnetic flux is approximately conserved, implying
\begin{equation}
B\propto a^{-2},
\qquad
\rho_B\propto a^{-4}.
\end{equation}
The present-day magnetic field is therefore
\begin{equation}
B_0
=
\left(\frac{a_{\rm rec}}{a_0}\right)^2
B_{\rm rec}.
\end{equation}

Using the recombination-era estimate, one finds
\begin{equation}
B_0
\approx
\frac{2}{\pi}
\sqrt{
\chi_B\,\mathscr S_{\rm peak}\,
T_0\,k_{0,\rm peak}^3
}.
\end{equation}

Expressed in terms of the present-day wavelength
$\lambda_0=2\pi/k_0$, this becomes
\begin{equation}
B_0
\approx
7.8\times10^{-46}\ {\rm G}\,
\sqrt{\chi_B\,\mathscr S_{\rm peak}}
\left(
\frac{30\ {\rm Mpc}}{\lambda_{0,\rm peak}}
\right)^{3/2}.
\end{equation}

For a representative peak wavelength
\(\lambda_{0,\rm peak}\sim 15\,{\rm Mpc}\), a typical peak squeezing
\(\mathscr S_{\rm peak}\sim 10^{-4}\), and $\chi_B \sim 1$, one finds
\begin{equation}
B_{0,\rm peak}
\sim
2\times10^{-48}\ {\rm G}.
\end{equation}
The smallness of the field is primarily due to the small value of
$\mathscr S_{\rm peak}$.

\subsection{Interpretation}

The result highlights two key features of the mechanism.

First, the recombination transition naturally selects a large coherence
scale, consistent with the peak of the weighted spectrum
$k^3\mathscr S_k$ identified in Sec.~IV.

Second, the amplitude of the resulting magnetic field is extremely small
in the minimal recombination scenario. The main limitation is not the
late-time survival of the field, but the smallness of the frozen
squeezing parameter itself.

The predicted amplitude is far below current observational bounds on primordial magnetic fields~\cite{Caprini2004}.
Therefore, while the mechanism provides a concrete realization of a
frozen non-equilibrium electromagnetic relic on cosmological scales, it
does not by itself account for the observed cosmic magnetic fields.
Additional enhancement mechanisms or subsequent amplification processes
would be required for a fully realistic magnetogenesis scenario.

\section{Summary and Discussion}
\label{sec:discussion}
In this work, we have proposed a new mechanism for generating a relic
electromagnetic excitation during the recombination--decoupling
transition, based on the rate-dependent thermodynamics of the cosmic
photon gas. Treating the photon sector as an open system coupled to the
electron plasma, we showed that a finite Thomson relaxation rate drives
a departure from instantaneous thermal equilibrium, leading to
non-adiabatic mode squeezing. As the relaxation rate rapidly decreases
across recombination, this deviation can no longer grow and instead
freezes out at a small but finite value, leaving behind a frozen
non-equilibrium electromagnetic relic.

A central result is that the reduced evolution equation can be recast,
via a canonical transformation, into a forced oscillator with a smooth
effective potential. In this formulation, the recombination epoch is not
an independent dynamical instability but a narrow transition layer in
which adiabatic tracking temporarily breaks down and the freeze-out
amplitude is selected. This provides a transparent interpretation of the
origin of the squeezing and clarifies the role of the relaxation
envelope, as well as the connection between low-frequency,
high-frequency, and transition-layer regimes.

We then related the non-adiabatic excitation to a magnetic relic
spectrum. The peak of the magnetic spectrum is controlled not simply by
the squeezing amplitude $\mathscr S_k$, but by the weighted combination
$k^3 \mathscr S_k$, leading to a characteristic present-day scale of
order $10$--$20$\,Mpc. The mechanism therefore naturally selects a
cosmologically large coherence scale associated with the recombination
transition.

At the same time, the minimal recombination-era realization considered
here predicts a very small present-day magnetic amplitude. The main
limitation is not the late-time survival of the relic, but the smallness
of the frozen squeezing itself. For this reason, the mechanism is more
naturally interpreted as a source of a frozen non-equilibrium
electromagnetic relic than as a complete explanation of the observed
cosmic magnetic fields.

The main contribution of this work is thus an analytic framework
connecting open-system non-adiabaticity, cosmological freeze-out, and
large-scale electromagnetic relic formation. In this sense, the
recombination-era scenario provides a concrete example of how a rapidly
varying thermal environment can leave a persistent imprint on
long-wavelength photon modes.

Several extensions are possible. The same mechanism may operate in other
cosmological transitions, such as reheating, where stronger
non-adiabatic excitation could in principle be generated, although its
survival may be limited by plasma damping. A plausible scenario is a
two-stage process in which an early-time excitation is subsequently
converted into a visible magnetic component during a later transition.
More generally, extensions to earlier epochs, including inflation, may
be possible in the presence of additional sources of non-adiabaticity or
effective dissipation. Exploring such scenarios requires a more complete
dynamical framework and is left for future work.

Overall, the present analysis establishes a new conceptual route from
the breakdown of thermal tracking to the generation of cosmological
electromagnetic relics, highlighting the role of open-system dynamics in
the thermal history of the Universe.

\begin{acknowledgments}
This work was supported by the National Research Foundation of Korea(NRF) grant with grant number RS-2023-00208047 and RS-2026-25483539 (H.K).
\end{acknowledgments}

\appendix

\section{Dissipative quasi-invariant formulation and squeezing parameter}
\label{ELR}

To define the squeezing parameter used in Sec.~III, we summarize a
dissipative quasi-invariant formulation in which the operator
$\hat{\mathcal I}(t)$ evolves under relaxation rather than remaining
strictly invariant.

We consider an operator obeying
\begin{equation}
\partial_t \hat{\mathcal I} - i[H,\hat{\mathcal I}]
= -\alpha(t)\bigl(\hat{\mathcal I}-g(t)\, H\bigr),
\end{equation}
where $\alpha(t)\ge0$ is the relaxation rate and
\begin{equation}
g(t)\equiv \frac{T_0}{T_e(t)}.
\end{equation}

\vspace{0.2cm}

For Gaussian states, we parametrize the density operator as
\begin{equation}
\rho(t)\propto
\exp\!\left[-\frac{\hat{\mathcal I}(t)}{T_0}\right],
\qquad
\hat{\mathcal I}(t)
=
\omega_{\mathcal I}(t)\!\left(\hat b^\dagger \hat b+\tfrac12\right),
\end{equation}
where $T_0$ is a fixed reference scale.

The quadratic operator can be written as
\begin{equation}
\hat{\mathcal I}
=
g_-\frac{\hat p^2}{2m}
+ g_0\frac{\hat p \hat x+\hat x \hat p}{2}
+ g_+\frac{m \hat x^2}{2},
\end{equation}
which defines the real functions $(g_-,g_0,g_+)$.

Substituting into the relaxation equation yields
\begin{equation}
\dot g_- = - 2g_0 -\alpha (g_- - g), \quad
\dot g_0 = \omega^2 g_- - g_+ - \alpha g_0, \quad
\dot g_+ = 2\omega^2 g_0 - \alpha(g_+- \omega^2 g ).
\end{equation}

\vspace{0.2cm}

The energy expectation value is
\begin{equation}
E
=\omega_{\rm eff}\,
\coth\!\left(\frac{\omega_{\mathcal I}}{2T_0}\right),
\end{equation}
with
\begin{equation}
\omega_{\rm eff}
= \frac{\omega_{\mathcal I}}{g}
\left(1+\frac{1}{\alpha}\frac{\dot \omega_{\mathcal I}}{\omega_{\mathcal I}}\right).
\end{equation}

This can be written as
\begin{equation}
\omega_{\rm eff} = \omega + \omega_{\mathcal I}\,\mathscr{S},
\end{equation}
which defines the squeezing parameter~\cite{Kim:2021wyp}
\begin{equation}
\mathscr{S}
=
\frac{g_0^2}{2\omega_{\mathcal I}^2 g_-}
+\frac12
\left(
\frac{1}{\sqrt{g_-}}-\frac{\omega}{\omega_{\mathcal I}}\sqrt{g_-}
\right)^2.
\end{equation}

A nonzero $\mathscr S$ therefore measures the departure from
instantaneous thermal equilibrium and quantifies the non-adiabatic
excitation of the mode.

\vspace{0.2cm}

Finally, the evolution of $\omega_{\mathcal I}$ can be written in the
relaxation form
\begin{equation} \label{domega}
\dot\omega_{\mathcal I}
=
\alpha(t)\big(g\,\omega_{\rm eff}-\omega_{\mathcal I}\big),
\end{equation}
showing that $\omega_{\mathcal I}$ evolves slowly once the relaxation
rate becomes small.

\section{Phenomenological model for $x_e(R)$ and thermal slip $g(R)$}
\label{App:xe}

We summarize the phenomenological inputs used for the ionization
fraction $x_e(R)$ and the thermal-slip parameter $g(R)$ that enter the
main analysis.

\subsection{Ionization fraction $x_e(R)$}

The recombination history is characterized by a rapid crossover from an
almost fully ionized state to a partially neutral plasma, followed by a
slow relaxation toward a residual ionization fraction. We model this
behavior by a smooth interpolation,
\begin{equation}
x_e(R)\simeq
x_e^{\rm tail}(R)
+
\bigl[x_e^{\rm pre}-x_e^{\rm tail}(R_*)\bigr]
\frac12\left[
1-\tanh\!\left(\frac{R-R_*}{\delta}\right)
\right],
\end{equation}
where $R_*$ denotes the center of the recombination transition and
$\delta$ its width.

Before the transition, the ionization fraction remains close to unity,
\begin{equation}
x_e^{\rm pre}\simeq 1.
\end{equation}

After decoupling, the evolution is governed by recombination without
photoionization, leading to a slow algebraic approach to a residual
value,
\begin{equation}
x_e(R)=x_{\rm res}+\frac{c_1}{R^{3/2}}+O(R^{-3}),
\end{equation}
with $x_{\rm res}\sim 10^{-4}$.

The width parameter $\delta$ is defined operationally by
\begin{equation}
\delta \equiv
\left|
\frac{x_e}{dx_e/dR}
\right|_{R=R_*},
\end{equation}
which measures the sharpness of the recombination layer. Typical values
are $\delta\sim 10^{-2}$, with a possible range extending to
$\delta\sim10^{-3}$ depending on the steepness of the transition.

This parameter plays a central role in the main text, as the peak
squeezing scale is determined by the matching condition
$Q(R_*)\,\delta^2\sim O(1)$.

For completeness, a simple representative model consistent with the
asymptotic behavior is
\begin{equation}
x_e^{\rm tail}(R)\simeq
\frac{1}{
1/x_{e{\rm dec}}
+\kappa\!\left(
1/t_{\rm dec}-\frac{1}{t_{\rm rec}R^{3/2}}
\right)},
\end{equation}
although the detailed functional form is not essential for the analytic
results, which depend only on the asymptotic scaling.

\subsection{Bath-branch coefficient $g(R)$}

In the main text, the reduced Gaussian state is parametrized as
\begin{equation}
\rho(t)\propto
\exp\!\left[-\frac{\hat{\mathcal I}(t)}{T_0}\right],
\end{equation}
where $T_0$ is a fixed reference temperature.
We choose $T_0$ such that, at the initial reference epoch, the thermal Gaussian state satisfies
\begin{equation}
\hat{\mathcal I}=\hat H .
\end{equation}
Equivalently, if the initial state is thermal with respect to the mode Hamiltonian,
\begin{equation}
\rho_{\rm ini}\propto e^{-\hat H/T_0},
\end{equation}
then the normalization of $\hat{\mathcal I}$ is fixed by this choice.

For an instantaneous electron bath temperature $T_e(R)$, the corresponding bath-selected thermal branch is
\begin{equation}
\rho^{\rm bath}(R)\propto e^{-\hat H/T_e(R)} .
\end{equation}
Expressed in the same Gaussian parametrization, this implies
\begin{equation}
\frac{\hat{\mathcal I}^{\rm bath}(R)}{T_0}
=
\frac{\hat H}{T_e(R)},
\end{equation}
or equivalently
\begin{equation}
\hat{\mathcal I}^{\rm bath}(R)=g(R)\,\hat H,
\qquad
g(R)\equiv \frac{T_0}{T_e(R)}.
\end{equation}

Thus, $g(R)$ should be interpreted as the dimensionless inverse-temperature coefficient of the instantaneous bath branch, measured relative to the fixed reference scale $T_0$. It does not represent the photon--electron temperature ratio itself. Rather, the actual departure of the photon mode from the bath branch is encoded dynamically in the deviation
\begin{equation}
\hat{\mathcal I}(R)-g(R)\hat H .
\end{equation}

During the recombination epoch, the electron temperature remains very close to the photon temperature, and both scale approximately as $a^{-1}$ while Compton coupling is efficient. Since $T_0$ is chosen at the reference epoch near recombination, one has
\begin{equation}
g(R)\simeq 1
\end{equation}
throughout the transition layer relevant for freeze-out, up to small corrections associated with the finite thermal evolution of the bath.

At later times, after full thermal decoupling, the electron temperature no longer tracks the radiation temperature and instead cools approximately as $T_e\propto a^{-2}$. In that regime one finds
\begin{equation}
g(R)\propto a \propto R.
\end{equation}
However, this late-time behavior does not play an important role in the recombination freeze-out dynamics studied in the present work, because the dominant non-adiabatic excitation is generated within the narrow transition layer where the relaxation rate decreases rapidly while $g(R)$ remains close to unity.

Therefore, in the minimal recombination scenario analyzed here, $g(R)$ may be treated as a slowly varying bath-branch coefficient, with the leading non-adiabatic effect arising primarily from the rapid decrease of the relaxation rate $\lambda(R)$ rather than from a large variation of $g(R)$ itself.

\section{Asymptotic branches and initial-condition selection}
\label{App:ini-cond}

The physical solution is most naturally specified by its early-time
asymptotic behavior rather than by imposing initial conditions at a
finite value of $R$.

\vspace{0.2cm}

In the limit $R\to 0$, the relaxation rate dominates,
\begin{equation}
\lambda \equiv \frac{\alpha}{R H}
\propto \frac{x_e}{R^{5/2}}\to \infty,
\end{equation}
so that the system is forced onto the instantaneous equilibrium branch,
\begin{equation}
G^-_k(R)\to g(R),\qquad
G^0_k(R)\to 0,\qquad
G^+_k(R)\to \frac{K^2}{R^2}\,g(R).
\end{equation}

Since $g(R)\to 1$ at sufficiently early times, this fixes the initial
condition uniquely up to small corrections.

\vspace{0.3cm}

At finite $R$, one may formally define an instantaneous fixed-point
branch by setting the time derivatives to zero. However, this branch does
not represent a true attractor at late times.

Indeed, as $R\to\infty$,
\begin{equation}
\lambda \propto \frac{x_e(R)}{R^{5/2}}\to 0,
\end{equation}
so that the relaxation toward the fixed-point branch becomes
inefficient. As a result, the solution ceases to track the instantaneous
equilibrium configuration and instead freezes out with a residual
deviation.

\vspace{0.2cm}

Thus, the evolution is characterized by an early-time adiabatic tracking
phase followed by a breakdown of tracking and late-time freeze-out. The
initial condition relevant for the main text is therefore the
early-time asymptotic branch given above.

\section{Asymptotic regimes of the canonical equation}
\label{App:asymptotics}

We briefly summarize the asymptotic behavior of the canonical equation
\begin{equation}
y_k''+Q(R)\,y_k=S(R),
\end{equation}
in the low- and high-frequency regimes.

\subsection{Low-frequency regime}

When the effective potential $Q(R)$ is small and the mode varies slowly,
the dynamics is dominated by the source term. In this regime one obtains
the quasi-static approximation
\begin{equation}
y_k \simeq \frac{S(R)}{Q(R)},
\qquad
u_k \simeq \frac{F(R)}{Q(R)}.
\end{equation}

Thus the deviation from equilibrium is a forcing-induced lag rather than
an oscillatory response.

Depending on parameters, $Q(R)$ may become negative in part of this
regime. In that case the homogeneous equation becomes locally
hyperbolic, but this does not correspond to a physical instability,
since the observable variable $u_k$ remains dressed by the relaxation
envelope.

\subsection{High-frequency regime}

When $Q(R)>0$ and varies slowly, the solution admits a WKB form
\begin{equation}
y_k \sim Q^{-1/4}\exp\!\left(\pm i\int^R \sqrt{Q}\,dR\right).
\end{equation}
The physical deviation is then
\begin{equation}
u_k \sim
R^{1/4}\exp\!\left[-\int^R \lambda\,dR\right]
Q^{-1/4}
\exp\!\left(\pm i\int^R \sqrt{Q}\,dR\right).
\end{equation}

In this regime, the dynamics is dominated by oscillatory freeze-out,
while the source contribution is parametrically suppressed,
\begin{equation}
u_k^{(p)} \sim \frac{F(R)}{Q(R)}.
\end{equation}

\vspace{0.2cm}

These two regimes represent different asymptotic realizations of the
same canonical equation and provide the basis for the transition-layer
analysis in Sec.~IV.

\section{Integrated relaxation profile}
\label{App:int-lambda}

For analytic estimates, it is useful to consider the integrated
relaxation factor
\begin{equation}
\int^R \lambda(\rho)\,d\rho,
\end{equation}
which controls the envelope in
\begin{equation}
u_k(R)\propto
\exp\!\left[-\int^R \lambda(\rho)\,d\rho\right].
\end{equation}

Using the phenomenological form of $x_e(R)$, the relaxation function can
be decomposed into a transition contribution and a post-transition tail,
\begin{equation}
\lambda(R)\simeq
\Gamma_k \frac{1}{R^{3/2}}
\left[
x_e^{\rm tail}(R)
+
\frac{1-x_*}{2}
\left(1-\tanh\!\frac{R-R_*}{\delta}\right)
\right],
\end{equation}
where $\Gamma_k$ is a mode-dependent constant.

The integrated profile therefore separates as
\begin{equation}
\int^R \lambda(\rho)\,d\rho
\simeq
\Gamma_k I_{\rm tail}(R)
+
\Gamma_k I_{\rm tr}(R).
\end{equation}

\paragraph{Transition contribution}

For a narrow transition layer, the factor $R^{-3/2}$ varies slowly, and
one finds
\begin{equation}
I_{\rm tr}(R)
\sim
\frac{1-x_*}{2R_*^{3/2}}
\bigl[R-|R-R_*|\bigr].
\end{equation}
Thus the transition produces a finite imprint that saturates once
$R>R_*$.

\paragraph{Tail contribution}

At late times, the ionization fraction follows
\begin{equation}
x_e(R)=x_{\rm res}+\frac{c_1}{R^{3/2}}+\cdots,
\end{equation}
which implies
\begin{equation}
I_{\rm tail}(R)
\sim
\int^R \frac{d\rho}{\rho^{3/2}}
\sim R^{-1/2}.
\end{equation}
Thus the post-transition contribution grows slowly and governs the
long-time behavior of the relaxation envelope.

\vspace{0.2cm}

\paragraph{Summary}

The integrated relaxation factor naturally separates into two
contributions:
\begin{itemize}
\item a finite transition imprint generated across the recombination
layer,
\item a slowly varying tail that controls the late-time evolution.
\end{itemize}

This structure underlies the freeze-out behavior discussed in Sec.~IV.



\end{document}